# Aerodynamic Design Considerations for Biconic Supersonic Air Intakes Revisited

J. P. S. Sandhu,[a1]  M. Bhardwaj,[a] N. Ananthkrishnan,[a] A. Sharma,[a] I. S. Park,[b] S. Jin[b] and J. H. Ryu[b]

[a]Yanxiki Tech, 152 Clover Parkview, Koregaon Park, Pune, 411001, India

[b]Agency for Defense Development, Daejeon, 305-600, South Korea

**Traditional design principles for determining the optimal intake ramp or cone angles, for ensuring no flow spillage at the intake cowl under design conditions, and for the form of the terminal shock in the intake duct are revisited. We show that it is preferable to select the ramp or cone angles to be somewhat smaller than that suggested by the Oswatitsch criterion. An offset cowl lip that slightly violates the shock-on-lip condition is found to be beneficial; in fact, an offset cowl can be arranged for conical intakes with no flow spillage at the cowl lip at all. Improvements to the total pressure recovery are seen when the terminal normal shock is replaced by a strong form of the oblique shock for two-dimensional ramp-type intakes, and with a Lambda shock in case of conical intakes. The necessary design modifications are simple and virtually cost-free. These results rewrite the ground rules for the aerodynamic design of supersonic intakes.**

## Nomenclature

$C_d$    =    intake pressure drag coefficient

$d$    =    diameter of rounded cowl lip

$h$    =    height of the intake

$\alpha$    =    shock angle with respect to the reference axis

---

[1] Senior R&D Engineer; jatinder@yanxiki.com, Corresponding author

| | | |
|---|---|---|
| $\beta$ | = | cowl lip geometric angle |
| $\Delta s$ | = | cowl lip offset distance along reference axis |
| $\delta$ | = | ramp or cone angle with respect to the reference axis |
| $\delta_4$ | = | cowl lip internal wedge angle |

*Subscripts*

| | | |
|---|---|---|
| $1,2,3$ | = | refers to first, second, third ramp or cone, respectively |

## I. Introduction

The design of supersonic air intakes is a fascinating exercise in multidisciplinary analysis and multi-objective optimization. From a purely aerodynamic perspective, the primary function of the air intake is to deliver the desired air mass flow to the engine under all flight conditions (Mach, altitude, incidence angles), in steady as well as maneuvering flight, with the best possible efficiency. Two measures of efficiency are commonly considered: 1. The total pressure recovery between the intake exit station and the free-stream flow, or any equivalent measure of energy efficiency, and 2. The additional drag that accrues due to the flow into and around the intake. The intake drag is usually taken to be comprised of additive drag due to the flow entering the intake, and cowl drag due to pressure and frictional forces acting on the external cowl surface. Aerodynamic forces acting on the internal cowl surfaces are normally included in the engine thrust calculation and do not form part of the intake drag accounting [1]. Additionally, for most engines, there is a constraint on the maximum flow distortion at the intake exit station that is permissible. The intake is also required to operate in a stable manner under all flight and engine operating conditions, which usually means that it should not be subject to buzz or unstart, where the terminal shock gets ejected out of the intake duct in either a steady or periodic manner. In either case, there is flow spillage at the cowl lip and the mass flow rate requirement of the engine can no longer be met, and there is a risk of engine flameout [2]. Besides these, there are several other factors that must be considered in the design of an air intake. These include, but are not limited to, the intake weight and the volume occupied by the intake, integration of the intake with the airframe, structural and thermal integrity of the intake and adjacent airframe components, the need for variable geometry features, the provision of bleed air, and the impact of the intake on the aircraft's stealth signature. An appreciation of many of these factors in the design and optimization of air intakes has been presented by Laruelle [3].

The basic principles of the aerodynamic design of supersonic intakes have remained essentially unchanged since the early days of the jet age [4]. These have been presented, following a traditional approach, in books on intake aerodynamics and design, such as those by Seddon and Goldsmith [5], and Mahoney [6], for example. Excellent presentations of the design of air intakes from the perspective of combat aircraft design are available in the books by Whitford [7] and Huenecke [8]. Broadly, three principles have generally been followed: First, the optimal choice of the intake ramp or cone angles to maximize the intake total pressure recovery according to the criterion by Oswatitsch [9]. Second, the requirement that all the ramp or cone shocks meet at the cowl lip at the design condition, referred to as the shock-on-lip condition. This ensures that the bounding surface of the capture stream-tube hits the cowl lip undeflected. Consequently, the entire mass flow within the capture stream-tube enters the intake duct with no spillage; this arrangement also yields zero additive drag. Third, the intake shock structure terminates in a normal shock designed to be located in the throat section of the intake duct. For external compression intakes, the throat is situated at the entrance to the intake duct, whereas mixed compression intakes have a throat section in the interior of the intake duct. In either case, the intake total pressure recovery improves as the normal shock is placed further upstream in the throat section. The precise location of the normal shock depends on the geometry of the intake duct and the back-pressure ratio, which is the ratio of the static pressures at the exit of the intake and the free-stream. At higher values of the back-pressure ratio, if the terminal normal shock is located too far forward in the throat, fluctuations in the flow due to atmospheric turbulence or variations in the combustor operating conditions may cause the terminal shock to be ejected from the intake duct resulting in buzz or unstart. Hence, a relatively safe location of the terminal shock in the intake throat needs to be selected, which is a compromise between the need to maximize total pressure recovery and the requirement to avoid unstable intake operation such as buzz or unstart [10].

While the above design principles broadly cover the requirements for intake performance (desired mass flow rate), efficiency (maximize total pressure recovery and minimize additive drag), and safety (avoid buzz or unstart), the question of minimizing the cowl drag is not explicitly addressed. The pressure drag is the dominant component of cowl drag and this arises from the increased pressure behind the coalesced intake shocks acting on an inclined cowl surface. It is easy to see that the requirement to minimize cowl pressure drag is in conflict with the aim of improving the total pressure recovery. By choosing intake ramp or cone angles lower than the optimum Oswatitsch values, one can create weaker shocks, hence the pressure behind the coalesced intake shocks will be lower, resulting in a reduced cowl pressure drag. Another option is to choose a smaller number of intake ramps (say, two ramps instead of three), which will allow the cowl to be placed at a shallower inclination angle, thereby reducing its projected frontal area, again causing lower cowl pressure drag. Either of these options sacrifices some total pressure recovery to gain an improvement in intake drag. Evaluating such a tradeoff between the multiple objectives of maximizing total pressure recovery while simultaneously minimizing intake drag with reasonable fidelity requires the use of a series of computational fluid dynamics simulations run in tandem with an optimization algorithm. Such an exercise has been carried out recently for the case of an internal compression intake by Brahmachary and Ogawa [11], and for an external compression intake with three ramp surfaces by Sandhu et al. [12]. These results show that it is possible to gain a significant improvement in intake drag at the cost of a marginal loss in total pressure recovery.

By and large, these design principles have been written with two-dimensional, ramp-type intakes in mind, but they have been applied, with some procedural changes, to axisymmetric, cone-type intakes as well [13]. It is well known that shock waves over conical surfaces differ from their two-dimensional, planar counterparts in a few significant ways [14]. Of particular interest is the fact that the streamlines behind a conical shock are not all parallel to the conical surface, and that the second conical shock in a biconic intake is not planar but curved. Consequently, solutions to the Taylor-Maccoll equation for conical flows are not as easily obtained as is the case for flows with planar oblique shocks [15]. Nevertheless, conical intakes — specifically, biconic intakes — have been widely used on a variety of aircraft and missiles. From the iconic conical intake of the MiG-21 and the semi-annular biconic intake of the Mirage-2000, to the intake of the Brahmos ramjet missile and the latest ramjet-powered artillery shell designs, the use of conical intakes is widespread and popular. It is, therefore, an apt juncture to re-examine the design principles for supersonic intakes as they apply to conical intakes in general, and to biconic intakes in particular. Several questions arise in this context. First, whether a tradeoff between total pressure recovery and intake drag, similar to that demonstrated in Refs. [11,12], is possible for biconic intakes as well. If so, then what is a reasonable method for choosing cone angles lower than the optimal Oswatitsch values. Second, whether the shock-on-lip condition is indeed necessary for conical intakes, and what is the penalty in terms of mass flow rate and additive drag in case this condition is violated. Third, whether the terminal normal shock can be modified or differently located in conical intake designs to gain an advantage in total pressure recovery or mass flow rate under off-design operating conditions. These questions are addressed in the present paper.

As may be expected, complex flows involving multiple shock interactions, expansion waves, and their interactions with the boundary layer, possibly including regions of separated flow, cannot be handled using analytical or approximate numerical methods. A full-fledged computational approach is necessary to analyze the flow over and through conical intakes, as illustrated in a recent book by Sandhu et al. [16]. The results presented in this paper are based on the numerical schemes described in Ref. [16] and references therein; those details are not reproduced here. Instead, we focus on addressing the three main questions raised above, first with reference to two-dimensional, ramp-type intakes, and then on extending those design considerations to biconic intakes.

## II.  Sub-optimal Oswatitsch Solutions

It is intuitively clear that smaller ramp angles create weaker ramp shocks which generate lower pressure behind the shock system, and hence induce a lower pressure drag on the external cowl surface. The net total pressure recovery (TPR) across the intake shock system is the product of the TPR across the individual ramp shocks and that across the terminal normal shock. In case of weaker ramp shocks and a stronger terminal normal shock, the TPR across the ramp shocks is higher but the TPR across the normal shock is poor, and hence the net TPR across the shock system is low. If the ramp angles are too large, the TPR across each ramp shock is low, and even though the TPR across the relatively weaker normal shock will be better, the net TPR will still be poor. Clearly, there is a sweet spot between these extremes where the maximum value of TPR is obtained; this is given by the Oswatitsch criterion [9]. According to this criterion, the optimum value of the ramp angles is such that the TPR across each ramp shock is identical.

**Table 1 Oswatitsch optimal solution for biconic intake at Mach 3 using Taylor-Maccoll theory**

| Shock | Upstream Mach number | Downstream Mach number | Incremental cone angle (deg) | Average flow turning angle (deg) | TPR corrected for curved shock | Net TPR |
|---|---|---|---|---|---|---|
| 1st cone | 3.0 | 2.35 | 18.5 | 11.0 | 0.953 | 0.953 |
| 2nd cone | 2.35 | 1.75 | 15.0 | 12.5 | 0.953 | 0.908 |
| Terminal normal | 1.75 | 0.63 | — | — | 0.833 | 0.756 |

Consider the case of a two-ramp planar intake designed as per the Oswatitsch criterion to give the optimal values of the ramp angles. For equal TPR across the two ramp shocks, the shock strengths must be identical. However, the upstream flow approaching the second ramp is necessarily at a lower Mach number than the free-stream Mach number, hence the flow at the second ramp must be deflected more than that at the first ramp for the two shocks to be of equal strength. In other words, the Oswatitsch criterion requires the incremental second ramp angle to be larger than the first ramp angle; this is a well known result [14]. Now consider a biconic intake at the same flow condition. The Oswatitsch criterion can be applied to the biconic intake by iteratively estimating the shock properties using the Taylor-Maccoll equation until convergence. The result from such a calculation for a biconic intake at Mach 3 is presented in Table 1. Note that the TPR across each of the conical shocks is identical as required by the Oswatitsch criterion; the net TPR calculated matches perfectly with that given in Ref. [14]. However, the incremental second cone angle of 15 deg in Table 1 is smaller than the optimal first cone angle of 18.5 deg. This occurs because the flow behind the first conical shock is not parallel to the wall at 18.5 deg; rather the average flow turning angle behind the first shock is only 11 deg. When this flow encounters the second cone at an incremental angle of 15 deg, the effective second cone angle is (18.5 - 11) + 15 = 22.5 deg, which is indeed larger than the first cone angle of 18.5 deg. The fact that the optimal geometric angle of the second cone in a biconic intake ought to be less than the first cone angle has not been well appreciated in the literature. This is one notable difference between the design of two-dimensional ramp-type intakes and axisymmetric cone-type intakes.

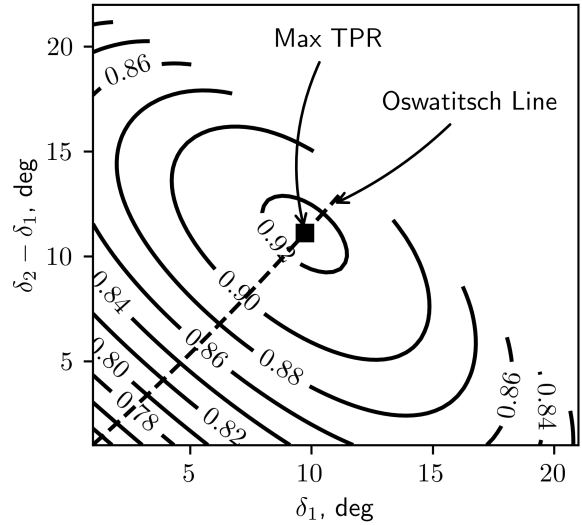

**Fig. 1 Theoretically calculated contours of net total pressure ratio for various values of the first ramp angle $\delta_1$ and the incremental second ramp angle $\delta_2 - \delta_1$ for a three-ramp planar intake ($\delta_3 = 33$ deg held constant).**

For a tradeoff between total pressure recovery (TPR) and intake drag (drag coefficient, $C_d$), a method of selecting ramp or cone angles smaller than the optimal Oswatitsch values is needed. This is explained with reference to Fig. 1 which shows iso-contours of TPR for a three-ramp rectangular intake for different values of the first and second ramp angles. The Oswatitsch optimum point giving the ideal maximum net TPR of 0.922 for the intake shock system is marked in Fig. 1 by a filled square. Several contours with lower values of net TPR are also indicated in Fig. 1. Clearly, for any given TPR contour, many solutions with different combinations of the first and second ramp angles are possible. Of these, only one point on each contour represents the case where the TPR across each ramp shock is identical — these are the sub-optimal Oswatitsch solutions. The locus of such points is marked by a dashed line in Fig. 1; of course, this includes the optimal Oswatitsch solution as well. An optimization scheme can work its way down this set of sub-optimal Oswatitsch solutions starting from the Oswatitsch-optimal TPR value, and trade off TPR for intake $C_d$. Such a study by Sandhu et al. [12] revealed that the multi-objective optimum solution yielded a 12% reduction in intake $C_d$ for a 0.2% loss in TPR. The corresponding sub-optimal Oswatitsch intake ramp angles were between 1 to 2 deg smaller than their optimal values.

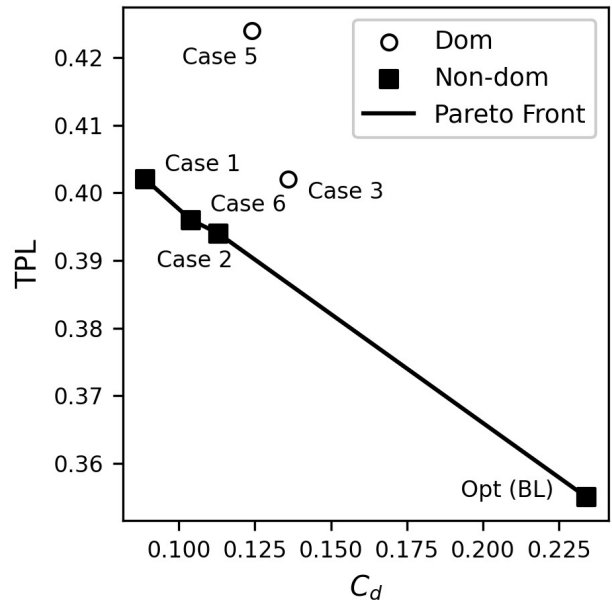

**Fig. 2  Tradeoff between total pressure loss TPL (1 - total pressure recovery) and intake drag coefficient $C_d$ for a biconic intake at Mach 3 using sub-optimal Oswatitsch solutions.**

Sub-optimal Oswatitsch solutions may be defined in a similar manner for conical intakes as well, though calculating them is a more laborious effort than for rectangular multi-ramp intakes, and used to trade off between TPR and intake $C_d$. Such an exercise has been undertaken by Sandhu et al. [17] for a biconic intake with the tradeoff indicated by the graph in Fig. 2, where total pressure loss, TPL = 1 - TPR, has been plotted. The Oswatitsch-optimal point, marked by 'Opt (BL),' shows the minimum TPL. The other points in Fig. 2 are sub-optimal Oswatitsch solutions with larger values of TPL but decreased intake $C_d$. It is seen that a reduction in intake drag of the order of 50% is achievable for a modest increase in total pressure loss. Of course, these numbers need to be converted into absolute values of airframe drag and engine thrust, but prima facie it appears conceivable that an increase in flight range could be achieved for, say, a ramjet-powered missile with a re-designed biconic intake using the concept of sub-optimal Oswatitsch solutions.

## III. Shock-on-lip Condition Violated

The shock-on-lip condition has always been one of the cornerstones of supersonic intake design. However, it has been recognized that there are circumstances where it is beneficial to slightly violate the shock-on-lip requirement [18]. Figure 3 shows two cases, one that satisfies the shock-on-lip condition, and the other that violates it slightly. In Fig. 3(a) with shock-on-lip, the free-stream flow encounters a compression surface at the cowl lip causing a strong shock to be formed there. If the cowl surface inclination is too large, a detached bow shock standing off slightly ahead of the cowl lip may be formed. This creates a flow spillage at the cowl lip, and the higher pressure behind the bow shock could induce a higher drag on the external cowl surface. In contrast, Fig. 3(b) shows the ramp shock deliberately missing the cowl lip. Instead, the flow over the final ramp passes over the cowl lip forming a weaker shock over the cowl. In this manner, the cowl drag may be reduced at the cost of a slight loss in mass flow rate entering the intake duct due to the spillage at the cowl lip. This concept has been numerically verified by Sandhu et al. [12] who found a notable reduction in intake drag coefficient for an insignificant loss in mass flow rate by displacing the cowl lip aft by as little as 1 per cent of the intake height.

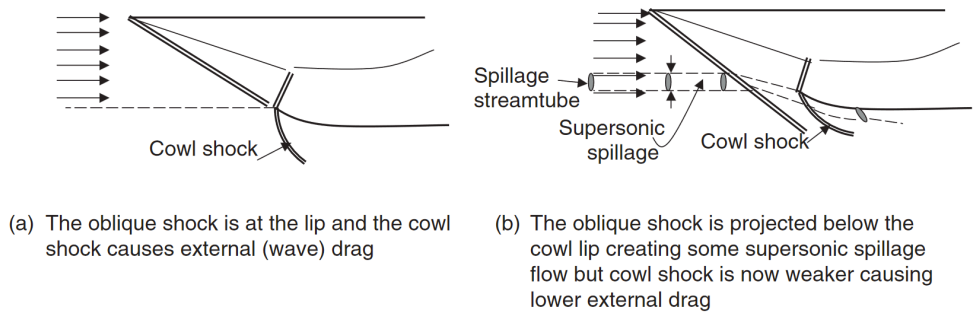

Fig. 3 Difference between the cowl shock in case of (a) shock-on-lip satisfied, and (b) shock-on-lip violated (taken from Farokhi [18]).

In case of rectangular ramp-type intakes, violation of the shock-on-lip condition by offsetting the cowl lip aft without altering the intake height necessarily implies flow spillage at the cowl lip, but that is not always the case for cone-type intakes — a fact that has not been widely recognized in the past. Figure 4 depicts a case of flow over a biconic intake where the cowl lip has been offset from the shock intersection point, but in such a manner that it lies on the bounding streamline of the capture stream tube. In this case, there is no flow spillage and the entire capture stream tube enters the intake duct. This is possible because streamlines behind a conical shock are curved and do not run parallel to the cone surface. The curved segment of the capture stream tube does contribute to additive drag, but this is exactly nullified by the reduced cowl drag due to the truncated cowl, as shown by Sandhu et al. [19] through numerical calculations. Hence, the shock-on-lip condition can be violated in this manner for conical intakes with no penalty on total pressure recovery, or intake mass flow rate, or intake drag. The advantage of offsetting the cowl as shown in Fig. 4 is realized under off-design operating conditions, such as flight at higher Mach numbers or at non-zero flow incidence angles. For small changes from the design condition, the shock intersection point does not enter the intake duct when the cowl lip is offset, as would be the case for a design with the shock-on-lip condition enforced. When the conical shocks enter the intake duct, they repeatedly reflect and push the terminal shock aft within the duct, and generally result in additional losses. However, when the cowl offset is excessive, the intake duct can only support a lower value of the back-pressure ratio before the critical position of the terminal normal shock is reached, and hence the total pressure recovery obtained is correspondingly lower [19].

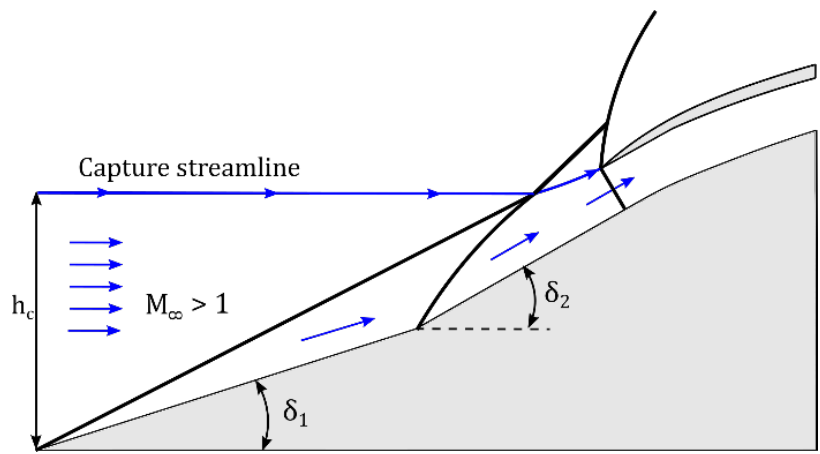

**Fig. 4  Biconic intake with cowl lip offset but aligned with the capture stream tube showing no flow spillage.**

## IV.  Modified Terminal Shock Structure

It is clear from Table 1 that the total pressure recovery (TPR) across the terminal normal shock is much worse than that due to the preceding conical shocks. The same is true in case of rectangular, ramp-type intakes as well. For mixed compression intakes, the offending shock is the oblique shock located at the cowl lip and directed into the duct — this is because the flow deflected at the upstream ramps is turned back through this single shock at the entry to the intake duct [20,21]. In case of sub-optimal Oswatitsch solutions, the pressure rise across the terminal normal shock is a little higher, and hence the TPR will be lower still. This provides the motivation to investigate possible alternative shock structures to replace the terminal normal shock.

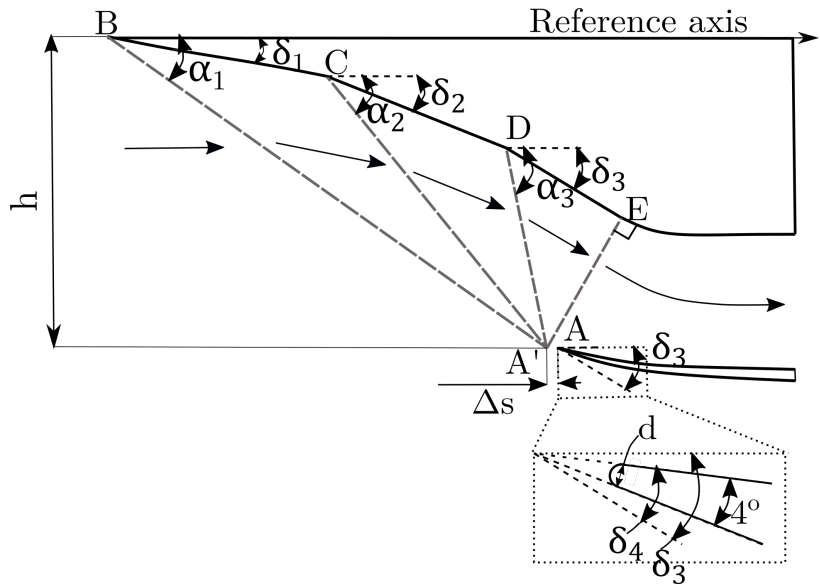

**Fig. 5  Three-ramp external compression intake with cowl lip axial offset $\Delta s$ and the effective cowl internal wedge angle $\delta_4$.**

One such alternative for rectangular, ramp-type intakes, put forward recently by Sandhu et al. [12], is to replace the terminal normal shock by a strong form of the oblique shock. This is achieved, with reference to Fig. 5, by rotating the cowl in a counter-clockwise sense through an angle $\delta_4$. The angle $\delta_4$ then forms an effective internal wedge angle at the cowl lip, so that the flow coming off the final intake ramp encounters a compression surface and has to deflect through an oblique shock to enter the intake duct. Of course, too large an angle $\delta_4$ will cause the shock to be detached and stand off from the cowl lip defeating the entire purpose. Numerical simulations by Sandhu et al. [12] have established that for small values of the internal wedge angle $\delta_4$, a strong form of the oblique shock may indeed be expected at the cowl lip. The flow behind the strong form of the oblique shock is subsonic and the TPR better than that of an equivalent normal shock. Further, the rotation of the cowl by an angle $\delta_4$ as depicted in Fig. 5 means that the cowl external surface is less inclined to the oncoming flow by the same angle $\delta_4$; hence, the pressure drag on the external cowl surface is reduced. Additionally, the effective wedge angle forces the oblique shock to be anchored at the cowl lip for a range of back-pressure values, so there is less degradation in TPR with falling back-pressure ratio near the critical region of the intake characteristic. Thus, this design fix appears to be advantageous on several fronts.

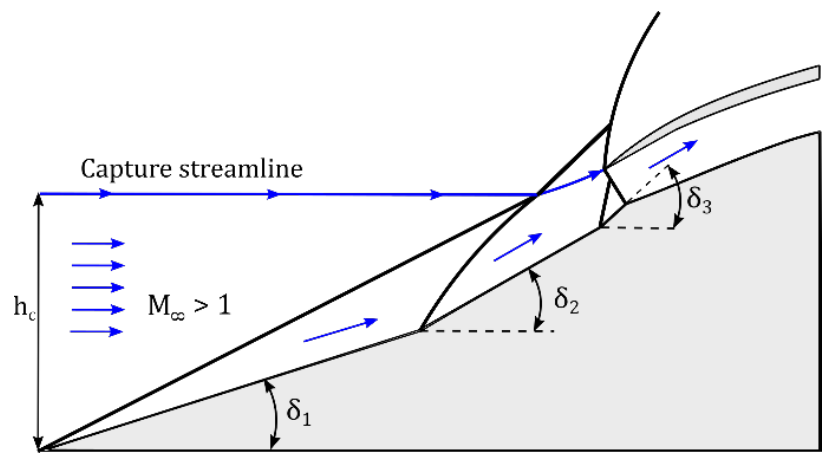

Fig. 6  Terminal Lambda shock structure in a biconic intake with added conical flare of angle $\delta_3$.

Unfortunately, the strong form of the oblique shock does not exist for conical flows. So a different device is required to mitigate the total pressure loss across the terminal shock in case of conical intakes. A simple, but extremely effective, solution to this problem has been presented recently by Sandhu et al. [22]. The fix is to introduce a conical flare of angle $\delta_3$, as marked in Fig. 6, such that the conical shock from the flare is designed to hit the inside of the intake duct just beyond the cowl lip. This shock typically intersects the terminal normal shock forming a Lambda shock structure, as sketched in Fig. 6. The exact point of intersection depends on the back-pressure which can relocate the normal shock slightly within the duct. The addition of the conical flare in this manner has no impact either on the flow upstream of the flare shock or on the pressure over the external cowl surface; hence, there is no effect on the cowl drag.

Sandhu et al. [22] have shown that the introduction of the Lambda shock structure instead of a solitary terminal normal shock improves the intake total pressure recovery by approximately 15% in case of a biconic intake at Mach 3. The behavior of the Lambda shock terminal structure under off-design conditions and its stability to perturbations in the flow remain to be investigated.

## V.  Conclusions

Several novel features for the aerodynamic design of supersonic intakes have been introduced in this work with a special focus on biconic intakes. First, the notion of sub-optimal Oswatitsch solutions is presented, which is relevant when both total pressure recovery and intake drag need to be simultaneously optimized. It is seen that by using ramp or cone angles somewhat smaller than the optimal values suggested by the Oswatitsch criterion, a significant reduction in intake drag is possible for a marginal loss in total pressure recovery. This applies to both rectangular ramp-type and axisymmetric cone-type intakes. Second, deliberately violating the shock-on-lip condition with an offset intake turns out to be a useful ploy in intake aerodynamic design. For conical intakes, this can be managed with no flow spillage at the design condition, whereas for rectangular, ramp-type intakes, a negligibly small loss in mass flow rate yields a notable decrease in cowl pressure drag. Third, considerable improvement in intake total pressure recovery can be achieved by modifying the traditional terminal normal shock to an alternative form. For rectangular, ramp-type intakes, there is the option of creating a terminal oblique shock of the strong form by the simple technique of arranging an effective internal wedge angle at the cowl lip. In case of conical intakes, the addition of a conical flare ahead of the cowl replaces the terminal normal shock by a Lambda shock structure which proves beneficial in terms of total pressure recovery with no losses otherwise. Application of these novel concepts to biconic intakes in recent research has been highlighted in this paper.

...